# Switching local magnetization by electric-field-induced domain wall motion

H. Kakizakai[1†], F. Ando[1†], T. Koyama[2], K. Yamada[1], M. Kawaguchi[1], S. Kim[1],

K.-J. Kim[1], T. Moriyama[1], D. Chiba[2]*, and T. Ono[1]*

[1]*Institute for Chemical Research, Kyoto University, Gokasho, Uji, Kyoto, 611-0011, Japan.*

[2]*Department of Applied Physics, The University of Tokyo, Hongo 7-3-1, Bunkyo, Tokyo, 133-8656, Japan.*

[†]These authors contributed equally to this work

*Correspondence to: dchiba@ap.t.u-tokyo.ac.jp, ono@scl.kyoto-u.ac.jp

**Electric field effect on magnetism is an appealing technique for manipulating the magnetization at a low cost of energy. Here, we show that the local magnetization of the ultra-thin Co film can be switched by just applying a gate electric field without an assist of any external magnetic field or current flow. The local magnetization switching is explained by the nucleation and annihilation of the magnetic domain through the domain wall motion induced by the electric field. Our results lead to external field free and ultra-low energy spintronic applications.**

Electric field effects (EFEs) on magnetic material have been intensively investigated in viewpoints of both the fundamental interests and the application potentials especially focusing on the modulation of the magnetic properties such as the coercivity or magnetic anisotropy,[1-6] the magnetic moment,[7-9] and the Curie temperature.[9-12] Especially, the EFEs in metallic ultra-thin films enables to make a device operation at room temperature possible. Maruyama *et al.* demonstrated that the magnetic anisotropy of ultra-thin Fe films can be modulated by the electric field.[4] More interestingly, the reversible change in the magnetic phase between ferro- to paramagnetic via the Currie temperature modulation induced by electric field has been reported.[9,12] The electric field effect on these magnetic properties even assists the magnetic-field-driven domain wall (DW) motion.[13-16] These reports on the EFE have opened up a way of controlling and manipulating the magnetic properties as well as the magnetization direction of the metallic films. The EFE is generally proportional to the voltage involving with a tiny current flow. The total energy required to control the magnetization can be quite low compared with an Oersted field generation or a spin polarized current application. Therefore, the electric field induced magnetization switching is a promising candidate for the next generation of ultra-low energy magnetic storages. Although there have been demonstrated a coherent magnetization reversal induced by EFE,[17,18] there is no report which succeeded in inducing a DW motion by EFE. In this paper, we demonstrate that the local magnetization switching using a domain nucleation and the DW motion just by applying an electric field.

Fig. 1 shows a schematic illustration of the device structure and the measurement configuration we used in this experiment. A perpendicularly magnetized Ta(2.5 nm)/Pt(2.4 nm)/Co(0.4 nm)/MgO(2.0 nm) from the substrate side was prepared by rf-sputtering on a thermally oxidized Si substrate. The film was then patterned into a 30-μm-wide Hall bar structure by photolithography and Ar ion milling followed by a deposition of the $HfO_2$(50 nm) insulator layer on the entire surface by the atomic layer deposition system. A transparent gate electrode of InSnO (ITO) was deposited by rf-sputtering on the top of the region of interest. A gate voltage $V_G$ was applied between the gate electrode and the Co layer. Here the positive $V_G$ was defined when the electron accumulation was induced in the Co layer side. Magnetic domain structures upon an application of the gate voltage were observed using a magneto-optical Kerr effect (MOKE) microscope with a polar-Kerr configuration. The Hall measurement is also performed to measure the magnetic hysteresis curve. The Hall resistance $R_{Hall}$ is generally proportional to the perpendicular component of the magnetization in the ferromagnetic layer because of the anomalous Hall effect. During the measurements, sample temperature was carefully controlled and monitored by a Peltier device and thermistor abutting on the sample. We found the Curie temperature $T_C$ of this sample was about 330K from the Hall measurements with various temperatures.

The loops shown in Figs. 2a and b show the Hall resistance $R_{Hall}$ curve obtained by sweeping out-of-plane magnetic field $\mu_0 H$ under $V_G$ of ±10 V at 315 K and 320 K, respectively. The coercivity and remanence clearly changed in response to the application of the gate voltage. Both the coercivity and the remanence increased with the positive $V_G$. The multi-domain state was expected to be formed from the non-zero remanence at 320 K. As displayed in Figs. 2c-e, the MOKE images taken at 320 K indeed showed a maze-shaped multi-domain structure at $\mu_0 H = 0$ T for $V_G = -10$, 0, and +10 V. In the images, the bright (dark) part indicated the area where the magnetization points upward (downward) direction. The domain structure was clearly changed upon application of $V_G$. We found that the positive $V_G$ increases the domain width $w$.

When a perpendicularly magnetized film is in a multi-domain state, $w$ is written in the following function,[19]

$$w \propto \exp\left(\frac{\sqrt{\varepsilon_{DW}}}{M_S^2 \cdot t}\right), \qquad (1)$$

where $\varepsilon_{DW}$, $M_s$, and $t$ are the DW energy per unit area, the saturation magnetization, and the magnetic film thickness, respectively. $\varepsilon_{DW}$ is proportional to $\sqrt{AK_u}$, where $A$ and $K_u$ are the exchange stiffness and the effective magnetic anisotropy energy, respectively. Hence, based on Eq. 1, $w$ should increase when $\varepsilon_{DW}$ increases and/or $M_s$ decreases. In our system, $M_s$ was expected to be increased with the positive $V_G$ from ref. 9 and saturated $R_{Hall}$ value in Fig. 2(b). Therefore, an increase of $\varepsilon_{DW}$ is deduced to be responsible for the increase of $w$, *i.e.*, the positive $V_G$ increases $\varepsilon_{DW}$. This is consistent with the previous experimental result or theoretical prediction of EFEs on $K_u$ and $A$ in Pt/Co system.[6,20] In our previous reports,[6] $K_u$ was found to increase by the positive $V_G$ in the vicinity of $T_C$. The increase of $A$ with the positive $V_G$ is most likely anticipated by the increase of $T_C$ in the previous reports.[9,12]. Thus, the increase of $\varepsilon_{DW}$ by the positive $V_G$ is expected to be a dominant factor for the increase of $w$.

As a result of the above electric field control mechanism, we demonstrated the switching of the local magnetization only by the gate voltage. The principle of the switching is based on the control of the domain nucleation and DW motion triggered by the application of $V_G$. Figure 3a shows a local area of the sample with $V_G = +10V$ at 315 K. At 315 K, as shown in Fig. 2a, the smaller remanence with $V_G = -10$ V indicates the multi-domain state, whereas the square hysteresis loop with $V_G = +10$ V indicates a uniform domain state, suggesting that the domain structure can be artificially recreated by $V_G$ at $\mu_0 H = 0$ T. The MOKE images confirms the domain structure recreation by $V_G$ at $\mu_0 H = 0$ T as shown in Figs. 3a-e. We started with applying a sufficiently large external magnetic field to the sample at $V_G = +10$ V in order to make the Co layer into single domain state (magnetization upward). We then turned off the magnetic field with keeping $V_G = +10$ V. The single domain state was kept as shown in Fig. 3a. By changing $V_G$ to -10 V, a domain with a downward magnetization was nucleated near the edge of the Hall bar (see Fig. 3b). As $V_G$ increase back to +10 V, the nucleated domain shrank, and eventually annihilated at $V_G = +10V$ (see Fig. 3c). The nucleation and annihilation of the downward magnetized domain was persistently and repeatedly observed with sweeping $V_G$ back and forth (Figs. 3d and e). The averaged and normalized intensity, thus the normalized magnetization, of the rectangular area highlighted by solid line in the images of Figs. 3a-e is plotted against the measurement duration in Fig. 4 together with the timing of $V_G$ application. It was obvious that the magnetization in this particular area is repeatedly switched after changing the sign of $V_G$. This means that the direction of the local magnetization is controllable just by applying $V_G$.

This electric field induced the domain nucleation and the DW motion observed here are interpreted as follows. First, the negative $V_G$ reduces $K_u$ or $A$, leading to a decrease of the $\varepsilon_{DW}$. As $V_G$ is further reduce to the negative, at some point, the multi-domain state is energetically favorable to reduce the magnetostatic energy of the system. The anisotropy energy at the edge of the wire is generally deteriorated during the sample fabrication *e.g.* by the ion milling process. Thus, the domain is expected to be more easily nucleated around the edge. When $V_G$ increases to the positive direction, the $\varepsilon_{DW}$ increases again and the system tries to shorten the length of the DW to reduce the total DW energy, consistent with the above argument of domain width $w$ as a function of $V_G$.

It is interesting to note that there is some amount of time delay between the $V_G$ application and the annihilation of the domain. This implies that $V_G$ actually modulates the DW velocity itself, supporting the previous reported that the DW velocity will be faster (slower) by applying a negative (positive) gate voltage.[14,15]

In conclusion, we realized the switching of the local magnetization only by the electric field application in the Pt/Co system. The local magnetization switching was performed by the nucleation and annihilation of the domain through the DW motion most likely due to the DW energy modulated by the electric field. We elucidated that the responsible factor for the DW energy modification is the response of $K_u$ and $A$ to the electric field. Our results present a guide way to achieve further efficient full electric field control of the magnetization direction, leading to ultra-low energy consumption spintronic devices.

**Acknowledgment**  This work was partly supported by JSPS KAKENHI (Grant Numbers 26103002, 25220604, 15H05702, 15H05419, and 2604316) and Collaborative Research Program of the Institute for Chemical Research, Kyoto University.

**References**

1) D. Chiba, M. Yamanouchi, F. Matsukura, and H. Ohno, Science **301**, 943 (2003).

2) M. Weisheit, S. Fähler Alain Marty, Y. Souche, C. Poinsignon, and D. Givord, Science **315**, 349 (2007).

3) D. Chiba, M. Sawicki, Y. Nishitani, Y. Nakatani, F. Matsukura, and H. Ohno, Nature **455**, 515 (2008).

4) T. Maruyama, Y. Shiota, T. Nozaki, K. Ohta, N. Toda, M. Mizuguchi, A. A. Tulapurkar, T. Shinjo, M. Shiraishi, S. Mizukami, Y. Ando, and Y. Suzuki: Nat. Nanotechnol. **4**, 158 (2009).

5) M. Endo, S. Kanai, S. Ikeda, F. Matsukura, and H. Ohno, Appl. Phys. Lett. **96**, 212503 (2010).

6) K. Yamada, H. Kakizakai, K. Shimamura, M, Kawaguchi, S. Fukami, N. Ishiwata, D. Chiba, and T. Ono, Appl. Phys. Express **6**, 073004 (2013).

7) M. Sawicki, D. Chiba, A. Korbecka, Y. Nishitani, J. A. Majewski, F. Matsukura, T. Dietl, and H. Ohno, Nat. Phys. **6**, 22 (2010).

8) M. Kawaguchi, K. Shimamura, S. Ono, S. Fukami F. Matsukura, H. Ohno, D. Chiba, and T. Ono, Appl. Phys. Express **5**, 063007 (2012).

9) K. Shimamura, D. Chiba, S. Ono, S. Fukami, N. Ishiwata, M. Kawaguchi, K. Kobayashi, and T. Ono, Appl. Phys. Lett. **100**, 122402 (2012).

10) H. Ohno, D. Chiba, F. Matsukura, T. Omiya, E. Abe, T. Dietl, Y. Ohno, and K.

Ohtani, Nature **408**, 944 (2000).

11) D. Chiba, F. Matsukura, and H. Ohno, Appl. Phys. Lett. **89**, 162505 (2006).

12) D. Chiba, S. Fukami, K. Shimamura, N. Ishiwata, K. Kobayashi, and T. Ono, Nat. Mater. **10**, 853 (2011).

13) M. Yamanouchi, D. Chiba, F. Matsukura, and H. Ohno, Jpn. Jour. Appl. Phys. **45**, 3854 (2006).

14) A. J. Schellekens, A. van den Brink, J. H. Franken, H. J. M. Swagten, and B. Koopmans, Nat. Commun. **3**, 847 (2012).

15) D. Chiba, M. Kawaguchi, S. Fukami, N. Ishiwata, K. Shimamura, K. Kobayashi, and T. Ono, Nat. Commun **3**, 888 (2012).

16) U. Bauer, S. Emori, and G. S. D. Beach, Appl. Phys. Lett. 101, 172403 (2012).

17) Y. Shiota, T. Nozaki, F. Bonell, S. Murakami, T. Shinjo, and Y. Suzuki, Nat. Mater. **11**, 39 (2012).

18) S. Kanai, M. Yamanouchi, S. Ikeda, Y. Nakatani, F. Matsukura, and H. Ohno, Appl. Phys. Lett. **101**, 122403 (2012).

19) B. Kaplan and G. A. Gehring: J. Magn. Magn. Mater, **128**, 111 (1993).

20) M. Oba, K. Nakamura, T. Akiyama, T. Ito, M. Weinert, and A. J. Freeman, Phys. Rev. Lett. **114**, 107202 (2015).

**Figure captions**

**Fig. 1**   Schematic illustration of device structure and the measurement setup.

**Fig. 2**   Magnetization curves under gate voltage $V_G = \pm 10$ V obtained by using the anomalous Hall effect at (a) 315 K and (b) 320 K. The MOKE microscope images taken under the application of (c) $V_G = -10$ V, (d) 0 V and (e) +10 V at 320 K.

**Fig. 3**   MOKE images at the edge of the film under the gate voltage (a), (c), (e) $V_G$ = +10 V, (b), (d) -10 V at 315 K. The observed area if the images is illustrated in the inset of (a). A domain is nucleated and annihilated repeatedly with applying gate voltage.

**Fig. 4**   The averaged and normalized magnetization at the rectangular area highlighted in the MOKE images of Figs. 3(a)-(e), and applied gate voltage $V_G$ plotted against measurement duration.

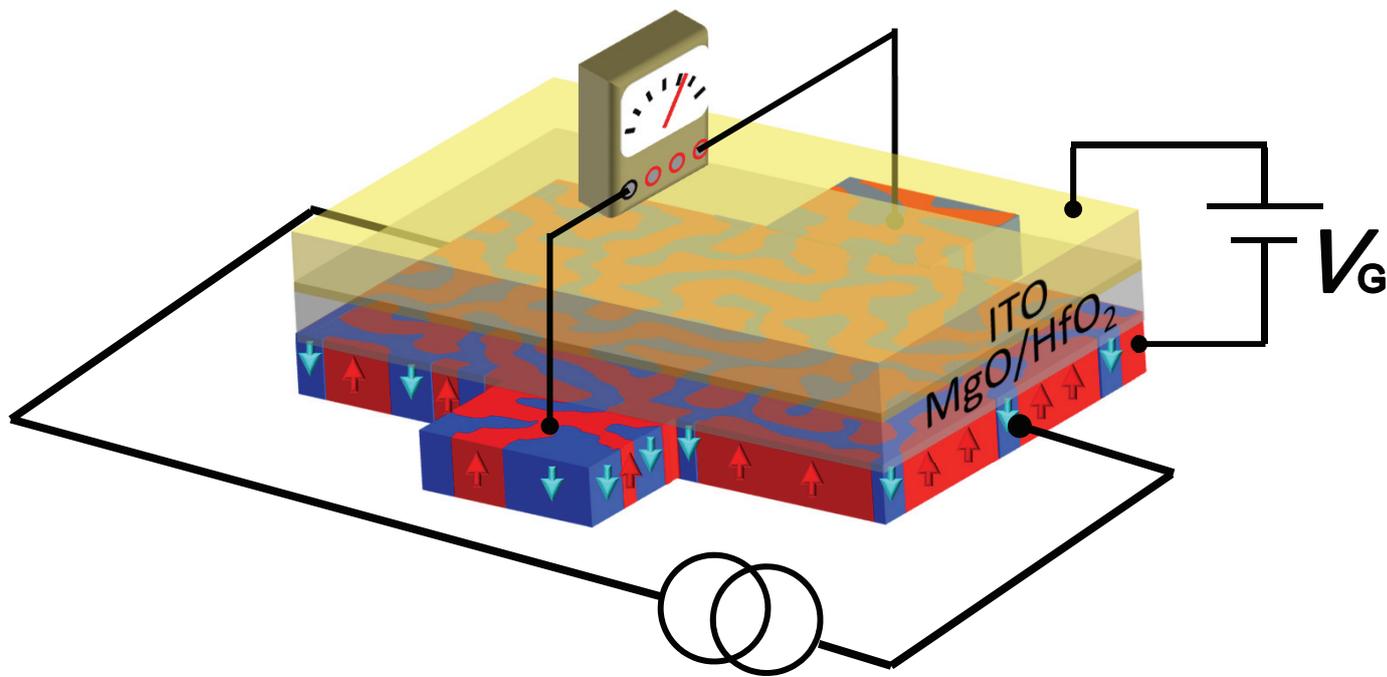

Kakizakai *et al.* Fig. 1

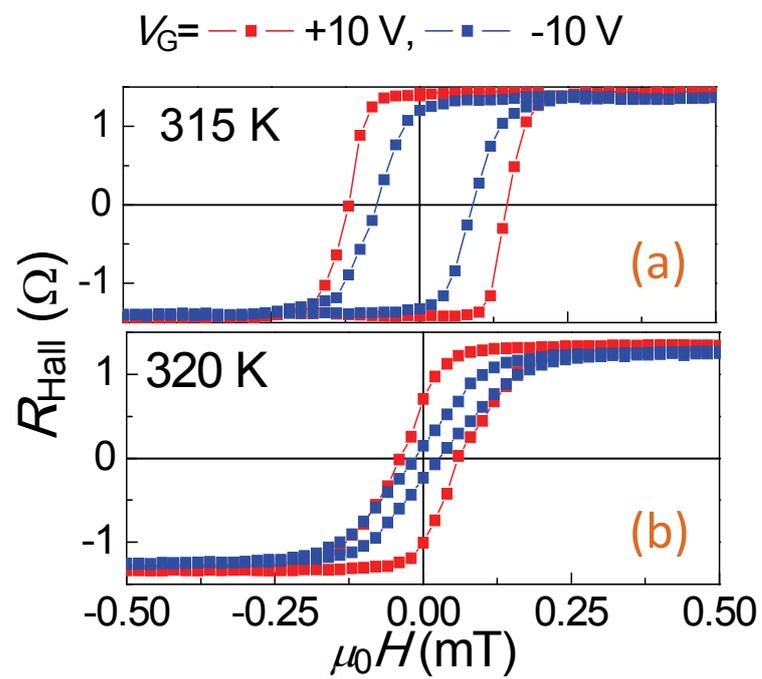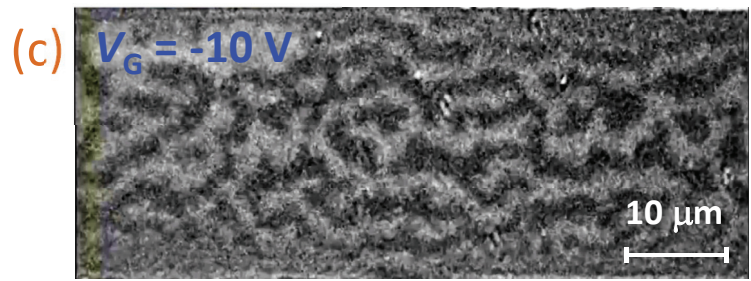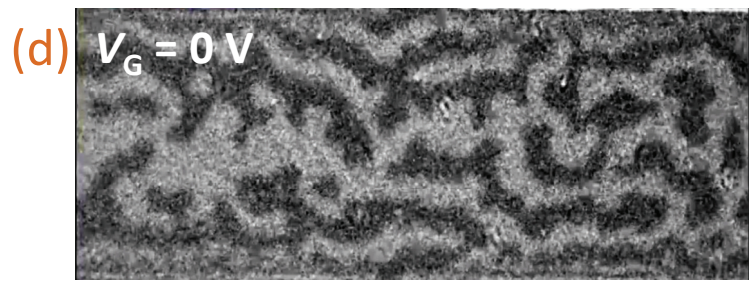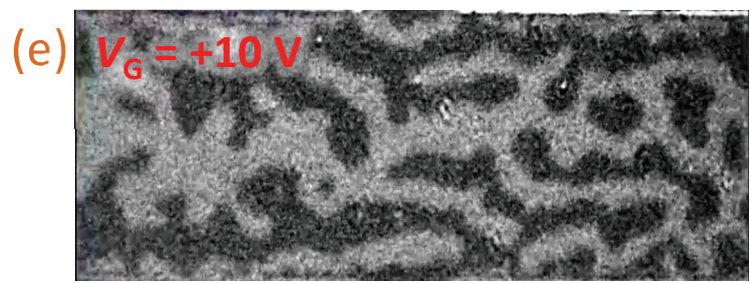

Kakizakai *et al.* Fig. 2

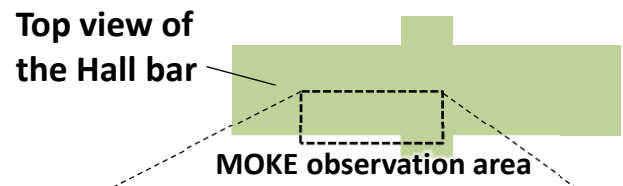
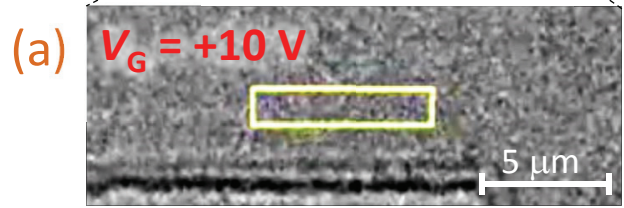
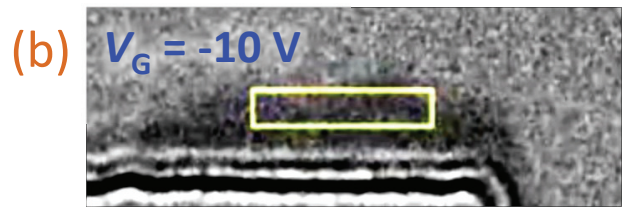
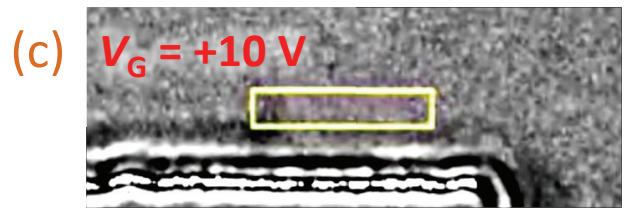
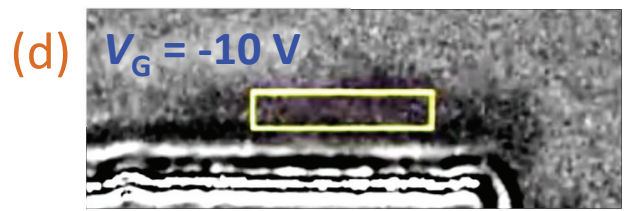
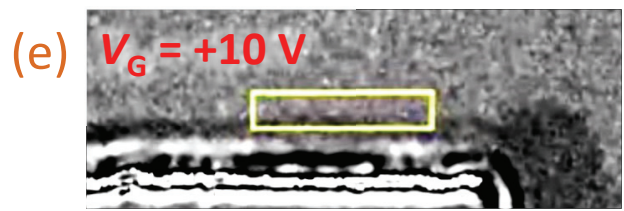

Kakizakai *et al.* Fig. 3

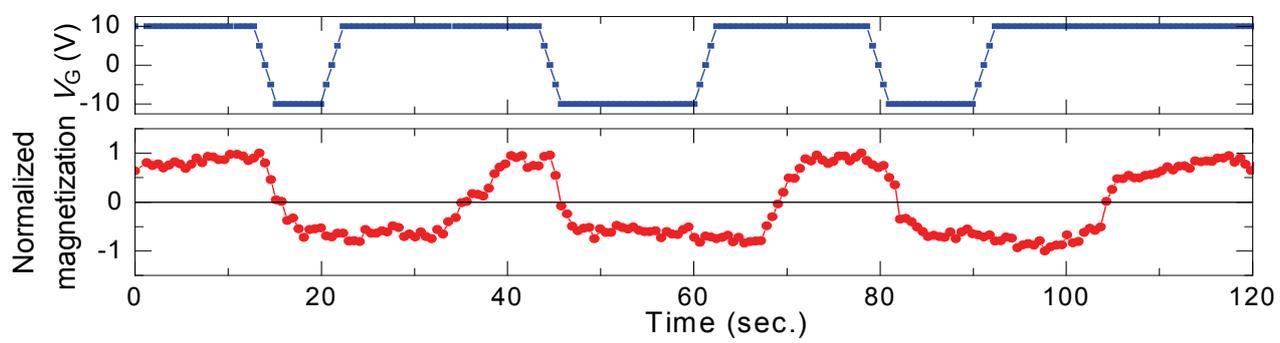

Kakizakai *et al.* Fig. 4